\documentstyle[12pt]{article}

\begin{document}

\begin{flushright}
\end{flushright} 

\vspace{2mm}

\vspace{2ex}

\begin{center}
{\large \bf  A simple derivation of} \\ 

\vspace{4ex}

{\large \bf the strong subadditivity inequalities} \\ 

\vspace{8ex}

{\large  S. Kalyana Rama}

\vspace{3ex}

Institute of Mathematical Sciences, C. I. T. Campus, 

Tharamani, CHENNAI 600 113, India. 

\vspace{1ex}

email: krama@imsc.res.in \\ 

\end{center}

\vspace{6ex}

\centerline{ABSTRACT}
\begin{quote} 

In this short note, we give a simple derivation of the strong
subadditivity inequalities: $S_{1 \; \cup \; 3} + S_{2 \; \cup \; 3}
\ge S_1 + S_2$ and $S_{1 \; \cup \; 2} + S_{2 \; \cup \; 3} \ge S_2 +
S_{1 \; \cup \; 2 \; \cup \; 3} \;$. The simplicity is due to the way
we represent the quantum systems.  We make a few remarks about such a
representation.

\end{quote}

\vspace{2ex}


\newpage


\vspace{4ex}





One considers quantum systems in pure states, two quantum
systems being entangled et cetera. Consider two quantum systems,
denoted as $1$ and $2$, which have zero initial entanglement
between them. Imagine building up entanglement between them by
the following protocol:

\begin{itemize} 

\item 

Take an EPR pair. Send one EPR partner to system $1$. This
partner will then interact with system $1$ and, together, they
will evolve unitarily.

\item

Similarly, the other EPR partner is sent to system $2$. This
partner will then interact with system $2$ and, together, they
will evolve unitarily. Systems $1$ and $2$ are otherwise kept
isolated.

\item 

Repeat the above steps $n_{12}$ times so that $n_{12}$ units of
entanglement entropy are built up between systems $1$ and $2
\;$.

\end{itemize} 

\vspace{2ex}

\centerline{\bf Two systems : subadditivity and triangle
inequalities }

\vspace{2ex}

Imagining that any entanglement entropies can be built up by
such a protocol, we represent systems $1$ and $2$ as
\begin{center}
$ \; \; 1 \; \; : \; \; \;$ 
\fbox{$ \; \; n_{12} \; ; \; \; N_1 \; ; \; \; \sim \sim \sim $}
\end{center}

\begin{center}
$ \; \; 2 \; \; : \; \; \;$ 
\fbox{$ \; \; n_{12} \; ; \; \; N_2 \; ; \; \; \sim \sim \sim $}
\end{center}
where $(n_{12}, \; N_1, \; N_2)$ are all non negative; $N_1$
denotes possible entanglement of system $1$ with some
environment; similarly $N_2$ denotes that of system $2 \;$; and
$\sim \sim \sim $ denotes the rest of the systems which have no
entropic content.

Keeping in mind the construction protocol for the entanglement
entropy, imagining operating it in reverse, and invoking the
monogamy of entanglement, it follows that the entropies $S_1$ of
system $1$, $S_2$ of system $2$, and $S_{1 \; \cup \; 2}$ of
systems $1$ and $2$ together are given by
\[
S_1 \; = \; n_{1 2} + N_1 \; \; ; \; \; \; 
S_2 \; = \; n_{1 2} + N_2 \; \; ; \; \; \; 
S_{1 \; \cup \; 2} \; = \; N_1 + N_2 \; \; .
\]
We immediately obtain the subadditivity and the triangle
inequalities \cite{nc}
\begin{eqnarray*}
S_1 + S_2 - S_{1 \; \cup \; 2} & = & 2 n_{1 2} \; \ge \; 0 \\
& & \\
S_{1 \; \cup \; 2} - \vert S_1 - S_2 \vert & = & 
N_1 + N_2 - \vert N_1 - N_2 \vert  \; \ge \; 0 \; \; . 
\end{eqnarray*}

\vspace{2ex}

\centerline{\bf Three systems : strong subadditivity
inequalities}

\vspace{2ex}

When three systems $1$, $2$, and $3$ are present, we represent
them as
\begin{center}
$ \; \; 1 \; \; : \; \; \;$ 
\fbox{$ \; \; n_{12} \; ; \; \; n_{1 3} \; ; \; \; N_1 \; ; 
\; \; \sim \sim \sim $}
\end{center}

\begin{center}
$ \; \; 2 \; \; : \; \; \;$ 
\fbox{$ \; \; n_{12} \; ; \; \; n_{2 3} \; ; \; \; N_2 \; ; 
\; \; \sim \sim \sim $}
\end{center}

\begin{center}
$ \; \; 3 \; \; : \; \; \;$ 
\fbox{$ \; \; n_{13} \; ; \; \; n_{2 3} \; ; \; \; N_3 \; ; 
\; \; \sim \sim \sim $}
\end{center}
where $(n_{1 2}, \; n_{1 3}, \; n_{2 3})$ and $(N_1, \; N_2,
N_3)$ are all non negative and their meanings are as before.
Again, as before, it follows that the various associated
entropies are given by
\begin{eqnarray*} 
S_1 & = & n_{1 2} + n_{1 3} + N_1 \\ & & \\
S_2 & = & n_{1 2} + n_{2 3} + N_2 \\ & & \\
S_3 & = & n_{1 3} + n_{2 3} + N_3 
\end{eqnarray*} 
and 
\begin{eqnarray*} 
S_{1 \; \cup \; 2} & = & n_{1 3} + n_{2 3} + N_1 + N_2 \\ & & \\
S_{2 \; \cup \; 3} & = & n_{1 2} + n_{1 3} + N_2 + N_3 \\ & & \\
S_{1 \; \cup \; 3} & = & n_{1 2} + n_{2 3} + N_1 + N_3 \\ & & \\
S_{1 \; \cup \; 2 \; \cup \; 3} & = & N_1 + N_2 + N_3 \; \; .
\end{eqnarray*}
We now immediately obtain the strong subadditivity inequalities
\cite{nc}
\begin{eqnarray*}
S_{1 \; \cup \; 3} + S_{2 \; \cup \; 3} - S_1 - S_2 & = & 
2 N_3 \; \ge \; 0 \\ & & \\
S_{1 \; \cup \; 2} + S_{2 \; \cup \; 3} 
- S_2 - S_{1 \; \cup \; 2 \; \cup \; 3} & = & 2 n_{1 3} 
\; \ge \; 0 \; \; .
\end{eqnarray*}

This is our simple derivation. Its simplicity and directness are
evidently due to the way we have represented the quantum
systems. To elevate this derivation to a proof, one has to show
that any quantum system can be represented by, or mapped to, the
representation given here -- atleast as far as the entropy
properties of the system are concerned. We do not know how to
show this.

However, considering our construction protocol for the
entanglement entropy and its operation in reverse, the
representation given here seems physically reasonable. As we
have seen, it leads to the subadditivity inequalities in a
simple and direct fashion. Therefore, at the least, such a
representation can be used to gain more intuition and, by
applying to multi systems, may also be used to obtain more
inequalities.

\vspace{3ex}

{\bf Acknowledgement:} 
We thank Samir Mathur for discussions. We also thank him for Nag
Memorial Lectures he delivered at our Institute in January 2014
where, among other things, he emphasised the importance of the
strong subadditivity inequalities.




\vspace{3ex}

\begin{thebibliography}{999}

\bibitem{nc}
Michael A. Nielsen and Isaac L. Chuang, \\
{\em Quantum Computation and Quantum Information}, \\
Cambridge University Press (2000). 

\end{thebibliography}
\end{document}